\documentclass[apj]{emulateapj}

\usepackage{graphicx}

\slugcomment{submitted to the Astrophysical Journal 17 October 2011}

\begin{document}

\title{Caught in the Act: the Assembly of Massive Cluster Galaxies at $z=1.62$}

\author{Jennifer M. Lotz\altaffilmark{1}, 
Casey Papovich\altaffilmark{2}, 
S. M. Faber \altaffilmark{3},
Henry C. Ferguson \altaffilmark{1},
Norman Grogin \altaffilmark{1},
Yicheng Guo \altaffilmark{4},
Dale Kocevski \altaffilmark{3}, 
Anton M. Koekemoer \altaffilmark{1},
Kyoung-Soo Lee \altaffilmark{5},
Daniel McIntosh \altaffilmark{6},
Ivelina Momcheva \altaffilmark{7},
Gregory Rudnick \altaffilmark{8},
Amelie Saintonge \altaffilmark{9, 10},
Kim-Vy Tran \altaffilmark{2, 11},
Arjen van der Wel \altaffilmark{12}, and
Christopher Willmer \altaffilmark{13}
 }
\altaffiltext{1}{Space Telescope Science Institute, 3700 San Martin Dr., Baltimore, MD 21218; lotz@stsci.edu}
\altaffiltext{2}{George P. and Cynthia Woods Mitchell Institute for Fundamental Physics and Astronomy, and 
Department of Physics and Astronomy, Texas A\&M University, College Station, TX, 77843-4242, USA}
\altaffiltext{3}{UCO/ Lick Observatory, Department of Astronomy \& Astrophysics, University of California,   Santa Cruz, CA 95064, USA}
\altaffiltext{4}{Department of Astronomy, University of Massachusetts, Amherst, MA, 01003, USA}
\altaffiltext{5}{Department of Astronomy, Yale University, New Haven, CT}
\altaffiltext{6}{Department of Physics, University of Missouri-Kansas City,  5110 Rockhill Rd., Kansas City, MO 64110, USA}
\altaffiltext{7}{Carnegie Observatories, Pasadena, CA, 91101, USA}
\altaffiltext{8}{Department of Physics and Astronomy, University of Kansas, Lawrence, KA, 66045-7582, USA}
\altaffiltext{9}{Max-Planck-Institut f\"{u}r Astrophysik,  Garching, Germany}
\altaffiltext{10}{Max-Planck-Institut f\"{u}r extraterrestrische Physik, Garching, Germany}
\altaffiltext{11}{Institute for Theoretical Physics, University of Z\"urich, CH-8057, Switzerland}
\altaffiltext{12}{Max-Planck-Institut f\"{u}r Astronomie,  Heidelberg, Germany}
\altaffiltext{13}{Steward Observatory, University of Arizona, Tucson, AZ, 85721, USA}

\begin{abstract}
We present the recent merger history of massive galaxies in a spectroscopically-confirmed
proto-cluster at z=1.62. Using {\it HST WFC3} near-infrared imaging from the
Cosmic Assembly Near-infrared Deep Extragalactic Legacy Survey (CANDELS), we select cluster galaxies
and $z \sim 1.6$ field galaxies with $M_{star} \ge 3 \times 10^{10} M_{\odot}$,  and determine
the frequency of double nuclei or close companions with projected separations less than 20 kpc co-moving
and stellar mass ratios between 1:1 and roughly 10:1. We find that four out of five spectroscopically-confirmed massive proto-cluster 
galaxies have double nuclei, and 42 $^{+13}_{-25}$ \% of all $M_{star} \ge 3 \times 10^{10} M_{\odot}$ cluster candidates
 are either in close pair systems or have double
nuclei.  In contrast, only 4.5 $\pm$ 2.6\% of the field galaxies are in close pair/double nuclei systems.  
The implied merger rate per massive galaxy in the proto-cluster is 3-10 times
higher than the merger rate of massive field galaxies at z $\sim$ 1.6, depending upon the assumed mass ratios.  
Close pairs in the cluster have minor merger stellar mass ratios ( $M_{primary}:M_{satellite}$ $\sim$ 6:1), while the field pairs
are typically major mergers with stellar mass ratios between 1:1 and 4:1. 
At least half of the cluster mergers are dissipationless, as indicated by their red 
colors and low 24 micron fluxes.  Two of the double-nucleated
cluster members have X-ray detected AGN with $L_x > 10^{43}$ erg s$^{-1}$, 
and are strong candidates for dual or offset super-massive black holes.  We conclude that the
massive $z = 1.62$ proto-cluster galaxies are undergoing accelerated assembly relative to the field population,  
and discuss the implications for galaxy evolution in proto-cluster environments. 
\end{abstract}


\section{Introduction}
The assembly of the most massive galaxies in the universe has long been a classical
problem for galaxy formation models.  Today these objects have $\ge$ $10^{12}$ $M_{\odot}$ of stars 
and live in the centers of galaxy clusters with $\ge$ $10^{14}$ $M_{\odot}$ dark matter
halos. Their stars are old with $\alpha$-element enhancements that point towards an intense
epoch of star-formation $>$10 Gyr ago (e.g. Thomas et al. 2005). However, their structures suggest a more
chaotic formation. The most massive galaxies have elliptical morphologies, often with
extended diffuse cD envelopes, boxy isophotes, and kinematics consistent with formation
via multiple gas-poor spheroid-spheroid mergers (e.g. Boylan-Kolchin et al. 2005, Khochfar \& Burkert 2003). 

Current hierarchical galaxy formation models invoke late assembly times for very massive galaxies 
via mergers of smaller galaxies who have already formed the bulk of their stars (de Lucia \& Blaziot 2007). 
Therefore a robust prediction of these models is that few exceptionally massive galaxies should 
exist in the early universe, and that their progenitors exist as many smaller sub-units at $z > 1$. 
In order to reconcile the difference between the star-formation histories and kinematic/
morphological structures, the majority of stars are formed in the progenitors prior to merging.   
The subsequent assembly of these progenitors at $z<1$ is expected to be largely dissipationless with little
associated star-formation.  

However, observations of the most massive galaxies at $z<1$ are not entirely consistent with this picture. 
Some direct look-back studies at $z < 1$ have found little evolution in the most massive early-type
galaxies. The number density of very bright ($> 2-4 L^*(z)$) early-type galaxies has not evolved significantly since $z \sim 1$  (Cimatti et al.  2006,  
Scarlata et al. 2007, Mancone et al. 2010).   The rest-frame luminosities and stellar masses of very bright distant galaxies suggest
that $\sim$ 80\% of the stars in today's most massive galaxies were already assembled by $z \sim 0.7$ (Brown et al. 2007).    
On the other hand, some studies find
stronger evolution in the number density of galaxies with stellar masses $> 10^{11} M_{\odot}$  (Ilbert et al. 2010; Brammer et al. 2011
but see Caputi et al. 2006).  Strong evolution is also observed for lower mass typical $L^*$ red galaxies,  which double in
number density and mass over the same epoch (e.g. Bell et al. 2004, Faber et al. 2007, Brown et al. 2007,  Ilbert et al. 2010, 
Brammer et al. 2011).    Direct evidence for mass growth by dissipationless merging is observed in  some very luminous 
red galaxies and the centers of $z < 1$ clusters  (e.g. Lauer 1988, Van Dokkum et al. 1999,  Tran et al. 2005,  White et al. 2007, 
Masjedi et al. 2008, McIntosh et al. 2008,  Brough et al. 2011),  and give the inferred mass growth rates of 1-20\% per Gyr. 

\begin{figure*}
\plotone{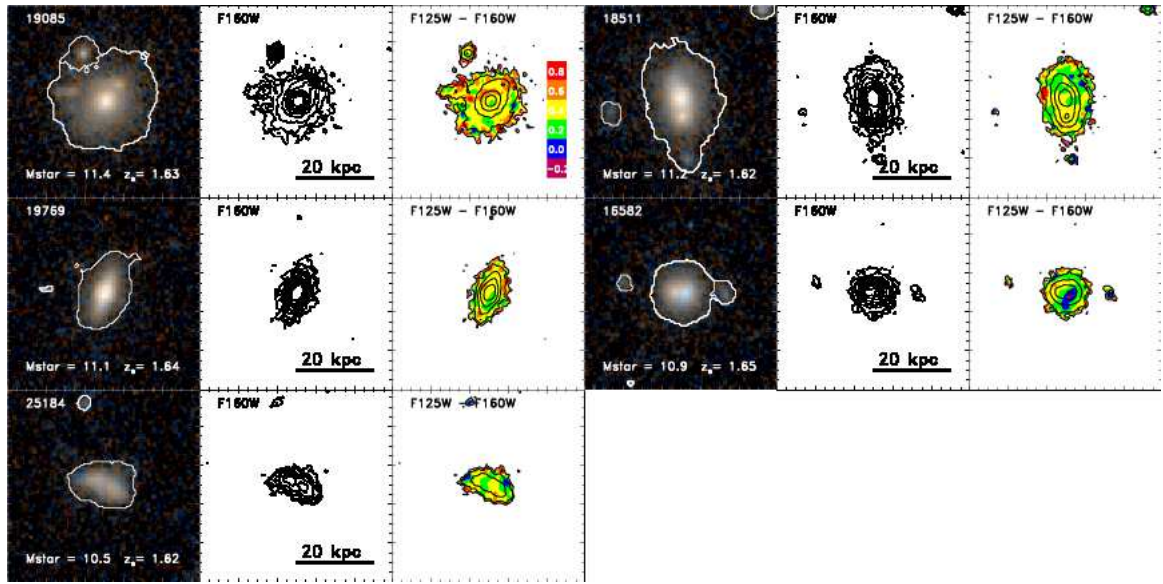}
\caption{The spectroscopically-confirmed $M_{star} > 3 \times 10^{10} M_{\odot}$ cluster members.  Left panels, RGB color maps from H/J+H/J images.
The F160W segmentation maps are shown as white contours. Center: H surface-brightness contours, at $\mu_H = 20.5 - 24$ in 0.5 mag per sq arcsec intervals. Right: J-H color maps where $\mu_H < 24.0$.  Four of the five massive cluster members have double nuclei/satellites separated by less than 20 co-moving kpc. The images are $6\arcsec \times 6\arcsec$ .}
\end{figure*}

\begin{deluxetable*}{clccccccc}
\setlength{\tabcolsep}{0.01in}
\tablewidth{0pc}
\tablecolumns{9}
\tabletypesize{\scriptsize}
\tablecaption{Spectroscopically-confirmed massive cluster galaxy properties}
\tablehead{ \colhead{CANDELS ID} & \colhead{$z_{spec}$}  & \colhead{$\alpha$}  & \colhead{$\delta$}  & \colhead{$H_{AB}$} & \colhead{log10[$M_{star}$]} & \colhead{$f_{24}$ \tablenotemark{3}} &\colhead{SFR \tablenotemark{4}} &\colhead{$L_{x}$[0.5-2 keV] \tablenotemark{5} } \\
\colhead{}  & \colhead{}  & \colhead{(J2000)}  & \colhead{ (J2000) }  & \colhead {} &\colhead{  ($M_{\odot}$)}  &\colhead{ ($\mu$Jy)}  &\colhead{  ($M_{\odot}$ yr$^{-1}$)} &\colhead{ ($10^{43}$ erg s$^{-1}$)} }
\startdata
  19085 \tablenotemark{6} & 1.634 \tablenotemark{1} & 34.58977 &  -5.17219 &  20.78  & 11.4 &   \nodata    &   \nodata  & \nodata\\
  18511 & 1.623 \tablenotemark{1}    & 34.58788 & -5.17583 &  20.55  & 11.2 &   \nodata   & \nodata   & $1.6 \pm 0.3$ \\
  19769 & 1.642 \tablenotemark{1}    & 34.57336 & -5.16781 &  21.05  & 11.1 &    \nodata      &     \nodata  &\nodata\\
  16582 & 1.649 \tablenotemark{1,2} & 34.57166 & -5.18490 &  21.57  & 10.9 &   79 $\pm$ 3  &  50 &  $2.7 \pm 0.4$  \\
  25184 & 1.622 \tablenotemark{2}    & 34.56322 & -5.13663 &  22.22  & 10.5 &    121 $\pm$ 4  &  86   &\nodata
\enddata
\tablenotetext{1}{Tanaka et al. 2010}
\tablenotetext{2}{Papovich et al. 2010}
\tablenotetext{3}{24 $\mu$m fluxes from Tran et al. (2010), SpUDS observations $^3$ ;  1-$\sigma$ detection limit $\sim$ 30 $\mu$Jy. }
\tablenotetext{4}{Star-formation rates are calculated from observed $f_{24}$ and spectroscopic redshifts,  assuming the redshift-dependent Rujopakarn et al. 2011 conversion of $f_{24}$ to $L(TIR)$.}.
\tablenotetext{5}{ Based on Chandra soft X-ray fluxes from Pierre et al. (2011).  Rest-frame 0.5-2 keV luminosities are calculated assuming the spectroscopic redshift and a power-law spectrum with index = 1.4.}
\tablenotetext{6}{Assumed central cluster galaxy} 
\end{deluxetable*}

Although clusters and associated dark matter halos continue to grow by accreting groups,  
galaxy-galaxy mergers may be suppressed in virialized clusters because of the high relative velocities of 
cluster members (typically 500-1000 km s$^{-1}$).  Matter recently accreted onto the central dark matter halo may be 
deposited into the satellite galaxy population and intracluster light, and not onto the central cluster galaxy (Brown 
et al. 2008; White et al. 2007; Conroy, Wechsler, \& Kravtsov 2007;  Gonzalez, Zaritsky, \& Zabludoff 2007; Rudnick et al. 2009). 
Galaxy groups with lower virial velocities (a few 100 km s$^{-1}$) are 
more conducive to galaxy mergers and the initial formation of massive red spheroidal galaxies than than rich clusters (e.g. Tran et al. 2008).  
The $z<1$ clusters which do show evidence of merging tend to be unrelaxed systems 
(e.g. MS1054: van Dokkum et al. 1999; Cl1604 : Kocevski et al. 2011).

The best way to disentangle the merger and star formation histories of massive cluster galaxies 
is to study their progenitors in overdense regions at $z>1.5$, i.e., in the environments expected to 
have the greatest galaxy merging and assembly
and during the period when the bulk of their stars formed.  If we want to catch
these galaxies in the act of star-formation,  these overdensities should not be biased towards galaxies which
have already formed the bulk of their stars.  Therefore the standard technique of selecting galaxy clusters via
an overdensity of red galaxies may miss newly forming systems at high redshift.
Likewise, if we want to catch these galaxies in the act of merging,  these overdensities should not be biased 
towards virialized relaxed systems whose internal velocity dispersions will suppress mergers.   Therefore
SZ and X-ray selected clusters may not be ideal places to study galaxy assembly, because their detection depends upon the presence of a halo of
hot virialized gas and hence selects already relaxed systems.   

In this paper, we study the recent merger history of massive galaxies in the spectroscopically confirmed
overdensity XMM-LSS J02182-05102 at $z=1.62$ (also known as IRC-0218A).   This proto-cluster was originally identified as a 20-sigma overdensity of 
high-redshift galaxies using an IRAC-color selection which identifies galaxies at $z>1.3$ regardless of 
spectral type (Papovich 2008). Spectroscopic follow-up by Papovich et al.(2010) and Tanaka et al. (2010)  confirmed
11 galaxies at $1.62 < z < 1.65$ within 1 projected physical Mpc of the central galaxy.   A marginal detection of diffuse X-ray emission is
associated with this proto-cluster  (Papovich et al. 2010; Tanaka et al. 2010),  and implies a virial mass $M_{200} = 7.7 \pm 3.8 \times 10^{13} M_{\odot}$ (Pierre et al. 2011).   However  the spatial structure of the overdensity indicates that it is not yet a virialized relaxed structure (Papovich et al. 2010). 
Likewise, the velocity dispersion is highly uncertain, with estimates ranging from 360 $\pm$ 90 km s$^{-1}$ (Pierre et al. 2011) to  860 $\pm$ 490 km s$^{-1}$
(Papovich et al. 2010).  The spectroscopic and photometric  redshift cluster members show evidence for a bright red sequence of galaxies (Papovich et al 2010, Tanaka et al. 2010),  as well as an excess of infrared luminous galaxies (Tran et al. 2010).  
Here we use high-spatial resolution {\it Hubble Space Telescope Wide Field Camera 3} ($HST$  $WFC3$) 
near-infrared F125W (J) and F160W (H) images from  Cosmic Assembly Near-infrared Deep Extragalactic Survey 
(CANDELS) pointing in the UKIDSS Deep  Survey (UDS) field to trace the merger history, structure, 
and resolved colors of the massive proto-cluster members.    The color-morphology and size-mass relations of the cluster members are
presented in a companion paper (Papovich et al. 2011).   Throughout this
 work, we assume $\Omega_m = 0.3$, $\Omega_{\Lambda} = 0.7$, and $H_0 = 70$ 
 km s$^{-1}$ Mpc$^{-1}$. 

\begin{figure*}
\plotone{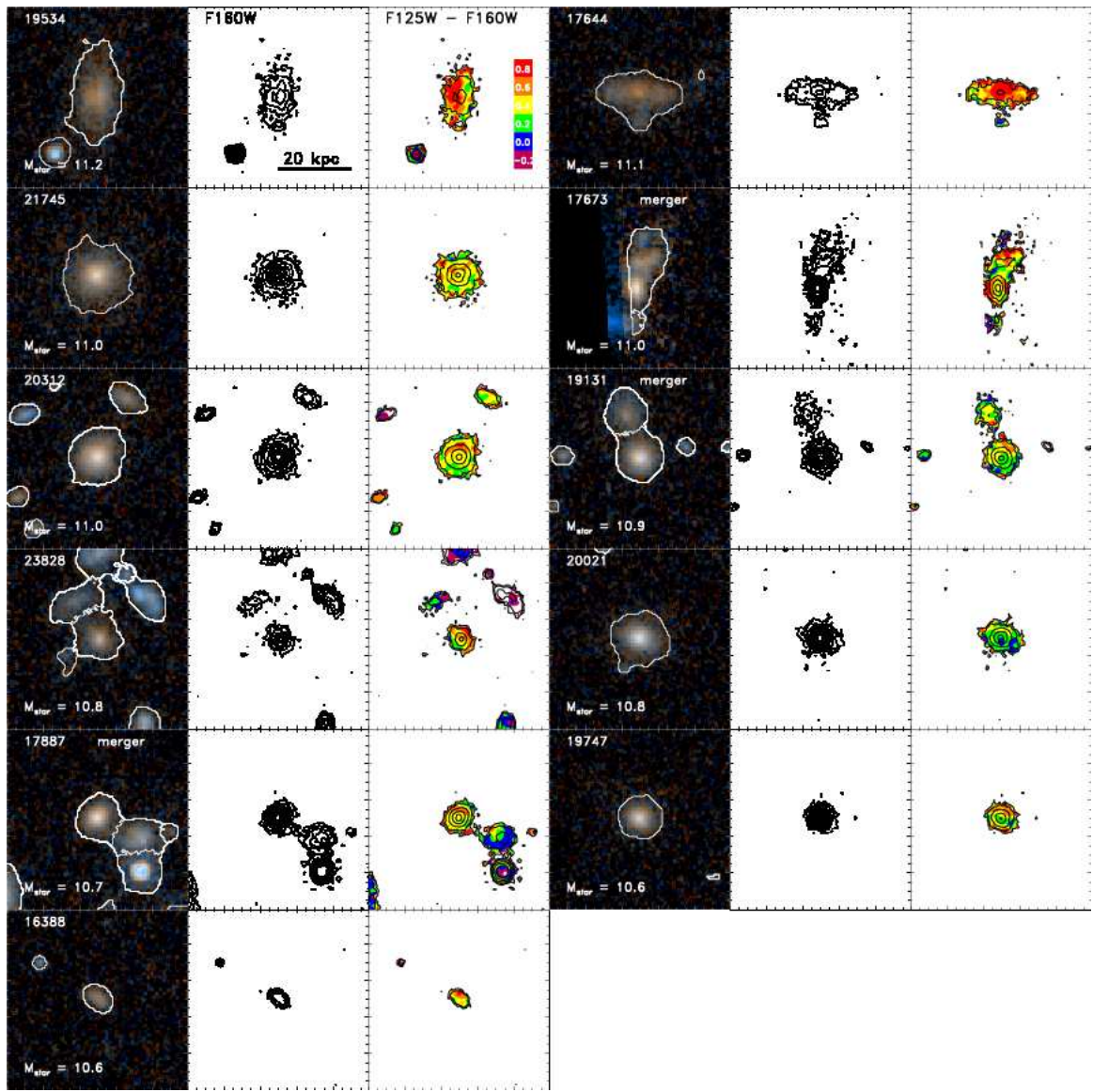}
\caption{The eleven additional photometric-redshift selected cluster candidates  ($M_{star} \ge 3 \times 10^{10} M_{\odot}$, $P(z) > 0.3$,  and projected distance from central cluster galaxy $< 1$ Mpc). Three photometric-redshift cluster candidates meet the merger criteria of a double nucleus or close companion brighter with $P_{zcl} > 0.3$  within 20 co-moving kpc. The image sizes, contours, and scalings are the same as Fig. 1.  }
\end{figure*}

\section{Observations}
The Cosmic Assembly Near-infrared Extragalactic Legacy Survey (CANDELS) is an $HST$ Multi-Cycle
Treasury Program (PI S. M. Faber and H. C. Ferguson).   CANDELS images five fields with the {\it WFC3-IR} camera (GOODS-N, GOODS-S, COSMOS, 
the Extended Groth Strip, and UKIDSS Deep Survey field) to 2-orbit depth total in F125W (J) and F160W (H), 
with deeper pointings in the GOODS-N and GOODS-S fields.  
The reduced combined images were drizzled to a 0.$\arcsec$06 pixel scale.
The details of the observational program are described in Grogin et al. (2011), and the 
details of the data reduction pipeline are described in Koekemoer et al. (2011). 
The UKIDSS Deep Survey field (UDS) was the first CANDELS field to be completed in January 2011. 

The UDS galaxies were detected and measured in the $HST$ F160W WFC3 images using SExtractor v2.5.0 (Bertin \& Arnouts 1996)
in combined `hot' and `cold' detection runs (Galametz et al., in prep.). The majority of the objects presented here were
detected in the `cold' run with the detection threshold set to 0.75$\sigma$, detection MINAREA = 5 pixels, 
and a 9$\times$9 pixel tophat convolution kernel. The combined F160W `hot + cold' segmentation map was then used as an input
template to perform multi-wavelength photometry on ground-based ($Subaru$ $BVRIz$,  $UKIRT$ $K$), $Spitzer$ (IRAC channels 1-4), and
other $HST$ (ACS F606W, F814W;  WFC3 F125W) images from the Subaru-XMM Deep Survey (Furusawa et al. 2008),  UKIRT IR Deep Sky Survey {\footnote[1]{http://www.nottingham.ac.uk/astronomy/UDS/data/dr3.html} (Almani et al., in prep; Williams et al. 2009),  Spitzer Extended Deep Survey {\footnote[2]{http://www.cfa.harvard.edu/SEDS} (PI. F. Giovanni), Spitzer UDS program {\footnote[3]{http:/irsa.ipac.caltech.edu/data/SPITXER/SpUDS}} (PI J. Dunlop) and CANDELS
 respectively,  using the TFIT software (Laidler et al. 2007).   
  
Photometric redshifts and stellar masses were calculated for the entire CANDELS-UDS field based 
upon the combined multi-wavelength photometry catalog. Photometric redshifts and probability distributions 
were computed using {\sc EAZY} code (Brammer et al. 2008).
Stellar masses were computed assuming Bruzual \& Charlot (2003) stellar population models,  
solar metallicities, a Chabier initial mass function, and Calzetti
et al. (2000) extinctions (see Papovich et al. 2001, 2006 for details). Based upon the 68\% confidence intervals of the redshift probability
distributions, the median uncertainty in the stellar masses is $\sim 20-30$\% at $z \sim 1.6$ and $M_{star} > 3 \times 10^{10} M_{\odot}$.  
We have compared the photometric redshifts to UDS spectroscopic redshift catalogs 
(Akiyama et al., in prep; Simpson et al. in prep; Smail et al. 2008,  Cooper et al. in prep) and find  $\delta z \sim 0.05 (1+z)$, with $\sim$ 6\% 
catastrophic outliers with $\delta z >  0.3 (1+z)$.

We follow Papovich et al. (2010, 2011) and
select cluster candidates using the integrated redshift probability 
\begin{equation}
\mathcal{P}_{zcl} \equiv \int^{0.05 (1 + z_{cl})}_{0.05 (1 - z_{cl})} P(z) dz
\end{equation}
where $z_{cl} = 1.625$.   Red cluster candidates typically have $\mathcal{P}_{zcl} > 0.5$. Spectroscopically-confirmed cluster members with
star-forming spectral energy distributions have less constrained photometric redshift probability distributions and  integrated $P_{zcl} \sim 0.3$. 
Therefore we require $P_{zcl} > 0.3$ for cluster candidacy in order to include both quiescent and star-forming galaxies.   
We will discuss field galaxy contamination in our photometric redshift cluster candidates in the following section. 

We have also matched the UDS samples to the $Spitzer MIPS$ 24 micron photometric catalog (Tran et al. 2010)
obtained by SpUDS$^3$. 
The 24 micron fluxes and spectroscopic or photometric redshifts were used to compute star-formation rates. 
At $z = 1.62$,  the observed 24 micron flux arises from rest-frame 9 $\micron$ emission,  which may be affected by
highly variable PAH emission (e.g. Smith et al. 2007).  We adopt the Rujopakarn et al. (2011) redshift-dependent conversions to 
extrapolate to a total infrared luminosity and corresponding star-formation rate.  Note that the Rujopakarn et al. (2011) 
conversion gives at factor of $\sim$ 2 lower star-formation rates for our sources 
than the Chary \& Elbaz (2001) templates.  The three-sigma detection limit of the 
$MIPS$ 24 micron catalog is $\sim 40 \mu$Jy or $\sim$ 20 $M_{\odot}$ yr$^{-1}$ at $z \sim 1.62$.   

The XMM-LSS J02182-05102 proto-cluster is serendipitously located in the edge of the CANDELS WFC3-IR imaging of the UDS field.
The cluster is only partially covered by $WFC3$ and has no $ACS$ parallel coverage (Grogin et al. 2011). 
Nevertheless, six out of 11 spectroscopically confirmed cluster galaxies (including the
central cluster galaxy) are located in the CANDELS observations.   Five of these objects have best-fit stellar masses greater than $3 \times 10^{10} 
M_{\sun}$ and are included in our analysis here (Table 1; Figure 1).  
We select 11 additional cluster member candidates with $P_{zcl} > 0.3$,  projected separation $<$ 1 Mpc from the central cluster galaxy, and
stellar masses $> 3 \times 10^{10} M_{\sun}$ (Figure 2).   Finally, we select a control sample of 134 field galaxies with $P_{zcl} > 0.3$,  
projected separation $>$  2 Mpc from the central cluster galaxy, and best-fit stellar masses $\ge 3 \times 10^{10} M_{\sun}$.    Based 
upon the surface density of the control sample,  we expect that roughly half of the photometric-redshift cluster candidates may be
background/foreground field galaxies.

\section{Merger rate in cluster v. field}
We identified merger candidates within our cluster and control samples as objects 
with double nuclei within the galaxy and/or close companions at a projected separation less than 20 kpc co-moving.   
While the vast majority of galaxies are cleaned detected and deblended in our HST F160W segmentation map, 
the distinction between galaxies with double nuclei and close pairs are particularly challenging for galaxy detection algorithms. 
In order to avoid incompleteness in our merger candidate samples, we visually inspected the F160W contour maps of all cluster and 
control sample galaxies for double nuclei within the SExtractor-defined segmentation maps (Fig 1 and 2, second panels) as well as 
searched for HST-detected companions within $\sim$ 20 kpc projected (2.4$\arcsec$) and $P_{zcl} > 0.3$.     We do not place
stellar mass constraints on the companions,  but note that the lowest mass companions have $M_{star} \sim 8 \times 10^{9} M_{\odot}$. 

We found that 4 out of the 5 spectroscopic cluster members have evidence of double/multiple nuclei in their F160W contour
maps (Table 1, Figure 1).   Out of the 11 additional photometric cluster candidates, 1 has multiple components (17673),  19131 is
in a 1:7 stellar-mass ratio close pair,  and 17887 is in a 1:6 stellar-mass ratio close pair. This is implies close pair fractions (double 
nuclei or close pair per descendant galaxy) of 80 $^{+13}_{-25}$\% for the spectroscopic cluster members and 44 $^{+13}_{-11}$ \% for all cluster members 
with stellar masses greater than $3 \times 10^{10}$ M$_{\odot}$.  (Uncertainties were computed assuming a binomial distribution with the {\sc R} statistical package {\footnote[4]{http://www.R-project.org}} routine binom.confint).  Four objects in control sample have double nuclei with the segmentation map. 
Eight of the control galaxies have close companions,  constituting 6 unique pairs. 
We derive a raw close pair fraction per descendant galaxy of 7.6 $\pm$ 2.4\% for our control sample.     We note that the lower measured pair fraction for
the combined spectroscopic and photometric redshift cluster candidates (44\%)  is consistent with the expectation that half of the photometric redshift cluster
candidates are field contaminants. 

We estimate the probability of another $P_{zcl} > 0.3$ UDS galaxy randomly falling within 20 kpc projected of our primary galaxy
sample ($P_{zcl} > 0.3$,  M$_{star} > 3 \times 10^{10} M_{\odot}$).  We  randomize the RA and DEC positions of all $P_{zcl} > 0.3$ CANDELS-UDS galaxies
within the CANDELS UDS field 1000 times.   The average fraction of false companions per primary galaxy is $2.7 \pm 1.3$\%.  Correcting
for this contamination gives a field pair fraction $ 4.9 \pm 2.7$\%.   Recent studies of field samples at $z \sim 2$ find pair fractions $\sim$ 5$-$18\% 
for projected separations $<$ 30 physical kpc $h^{-1}$ and stellar mass ratios between 1:1 and 10:1  (Williams et al. 2011; Man et al. 2011; Newman et al. 2011).   Our field pair fraction at $z \sim 1.6$ is consistent with these measurements, assuming the timescale for identifying close pairs at projected $<30$ physical kpc is roughly twice the timescale for finding pairs at $<20 $ co-moving (7 physical) kpc (e.g. Lotz et al. 2010). 

We also estimate the false companion fraction for the cluster galaxy sample.  The proto-cluster is an overdense region and false pairs may arise from galaxies associated with the cluster but are not interacting.   In order to maintain the cluster's density profile,  we randomly scattered the positions of all 67 cluster candidates ($P_{zcl} > 0.3$,  $<$ 1 Mpc from central cluster galaxy, $M_{star} > 8 \times 10^{9} M_{\odot}$) 
 to within 10\arcsec of their original positions 1000 times.   
We find that the average fraction of false companions per massive cluster galaxy is 1.6 $\pm$ 0.8\%.    Therefore we derive corrected pair fractions
78 $^{+13}_{-25}$ \% for the spectroscopic sample and 42 $^{+13}_{-11}$ for the combined spectroscopic and photometric redshift cluster samples. 

Assuming that the dynamical friction timescales and mass ratios for the cluster and control samples are similar, then
the implied cluster merger rate is more than ten times the field merger rate at $z = 1.6$.   
We argue that the sub-components and companions associated with each cluster galaxy  
are within close physical proximity and have a high probability of merging within 1-2 Gyr (by redshift $\sim$ 1).   
Our simulations suggest that small separations are unlikely to occur via random projections within the proto-cluster. 
Although the cluster is, by definition, an overdense region, we have selected objects within such small
separations that they also unlikely to be non-interacting cluster galaxies viewed in projection. 
The spatial distribution of galaxies and X-ray structure of the over-density suggest that it consists of several groups which have 
not yet formed a virialized cluster.  Therefore the relative velocities of galaxies in close proximity
are likely low enough for a mergers (i.e. a few hundred km s$^{-1}$ found in group environments) rather than a large relative velocities
( 500-1000 km  s$^{-1}$) found in massive cluster environments.   

We examine the stellar mass ratios of the close pairs as a function of the primary galaxy stellar mass in Figure 3.   
The two pairs in the cluster have stellar mass ratio $M_{primary}/M_{satellite}$ $\sim$ 6:1, consistent with minor mergers.
In contrast,  the field pair samples are more likely to have stellar mass ratios $\sim$ 1:1 - 4:1,  consistent with major majors. 
Our samples are too small to draw strong conclusions,  but suggest that minor mergers play a significant role in the
buildup of massive cluster galaxies.  Determining the stellar mass ratios of the multiple nuclei systems is hampered by the lack 
of high-resolution $HST$ data blueward of the 4000\AA\  break, and is beyond the scope of this paper.

The cluster mergers may have longer dynamical friction timescales than the typical field pairs if they have
systematically greater mass ratios and/or higher relative velocities.  
Simple dynamical friction arguments (Binney \& Tremaine 1987) imply that the decay time scales as $M_{primary} / M_{satellite}$.   
In the extreme case that all detected cluster mergers are minor mergers with typical 
merger timescales $\sim$ three times the field sample,  then number of mergers per massive cluster galaxy 
is roughly 3-4 times the field merger rate.
 
\begin{figure}
\plotone{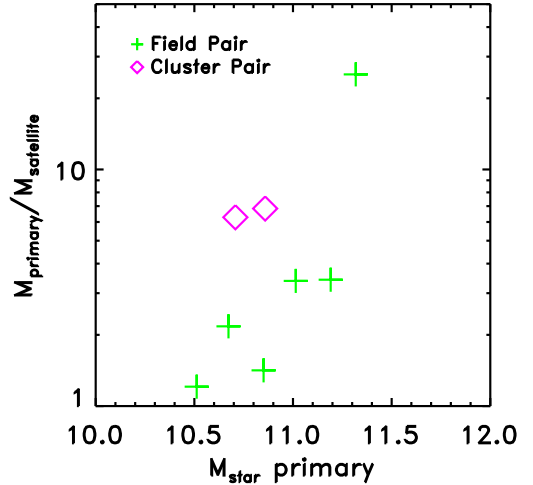}
\caption{The stellar mass ratio for our 2 cluster pairs and 6 field pairs as a function of primary galaxy stellar mass.}
\end{figure}

\begin{figure*}
\plotone{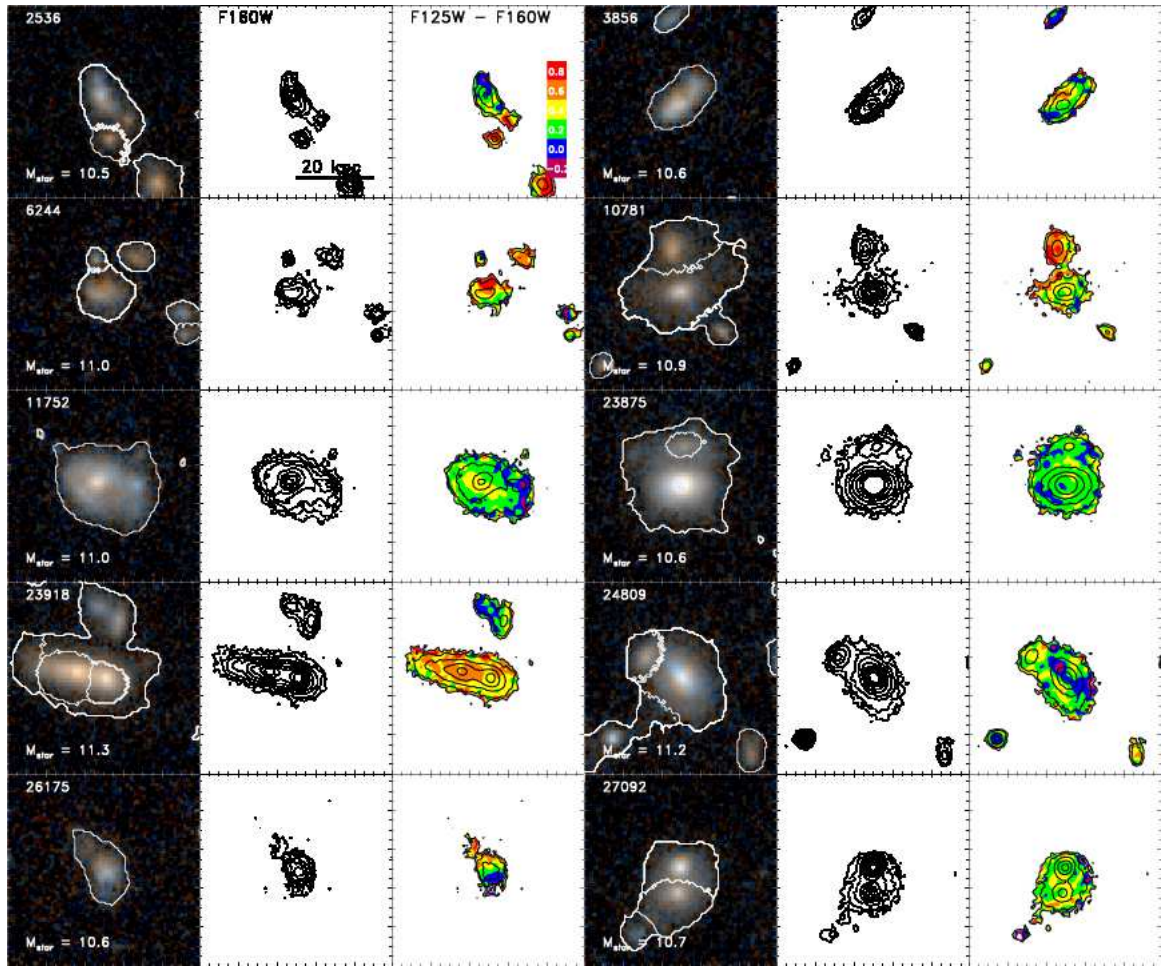}
\caption{The double nuclei/close pair candidates from the UDS control sample ($M_{star} > 3 \times 10^{10} M_{\odot}$, $P_{zcl} > 0.3$, and projected distance from central cluster galax $> 2$ Mpc).  Thirteen  UDS control objects show double nuclei or a close companion with $P_{zcl} > 0.3$ separated by less than 20 co-moving kpc,  resulting in 10 unique systems.  The white contours show the F160W segmentation map. Close pairs are counted once. }%
\end{figure*}

\section{Recent Star-Formation Histories of Mergers} 
We examine the $J-H$ color maps (Fig. 1 and 2) to constrain the relative colors of the merging pairs.  
Observed $J-H$ is roughly rest-frame $B-R$ (0.48- 0.61$\mu$m) at $z= 1.625$.  
We find that the two most massive cluster galaxies are consistent with red-red galaxy mergers.   
19085 is a massive red spheroidal galaxy (see Papovich et al. 2011) with 2 or 3 additional nuclei/companions of similar $J-H$ color
($\sim 0.4$). 18511 is a spheroid-dominated galaxy of $\sim$ 0.2 mag bluer in $J-H$ color than 19085 with a redder
second nucleus.  The less massive cluster merger 16582 has one red nucleus ($J-H = 0.4$) and one blue nucleus ($J-H = -0.2$).   
25184 has a double-peaked H-band surface brightness profile,  $J-H \sim 0.2-0.4$ colors, 
and a disky structure consistent with either a double nucleus or patchy dust.  None of the photometric redshift cluster merger
candidates have observed $J-H$ colors bluer than 0.2 for both components.    In contrast,  many of the field mergers are blue
with one or both components showing $J-H < 0.2$. 

Following Papovich et al. (2011), we also use the integrated $z-F125W$ and $F125W-$3.6$\mu$m colors to constrain whether the  galaxies 
are `quiescent' or `star-forming'.  At $z \sim 1.6$, these observed colors are similar to the rest-frame $U-V$ and $V-J$ colors used by 
Wuyts et al (2007) and  Williams et al. (2009) 
to distinguish between galaxies with relatively low star-formation rates and UV-bright/dusty galaxies with high star-formation rates.  
We find that two of seven cluster merger candidates are classified as star-forming while the remainder are classified 
as quiescent (left, Figure 5).  The $z-F125W$ and $F125W-$3.6$\mu$m colors are also shown for the field samples (right, Figure 5).  

The spatially-resolved and integrated colors of the cluster merger candidates suggest that most of the cluster mergers are not 
starburst systems.  However, some of these quiescent systems may be forming stars.  
We have matched the cluster candidates to the Spitzer MIPS 24 micron sources with 3\arcsec.   Seven cluster candidates have 3-sigma detections, 
including four mergers.    Forty-eight objects in the control sample are detected as 24 micron sources, including three merging systems.  
Assuming the 24 micron flux arises from star-formation rather than AGN activity, the cluster mergers
have star-formation rates $\sim$ 50-150  $M_{\odot}$ yr$^{-1}$. High star-formation rates can also be seen in the field galaxy samples. 
The two most massive cluster mergers are not detected in the MIPS 24 micron data.  One of the cluster 24 micron
sources (16582) is also a Chandra X-ray detection with an  AGN-like X-ray luminosity.  

In Figure 6,  we compare distribution of spectral type, stellar mass, and 24 micron-derived star-formation rates for
the cluster and control samples.  The mergers generally reflect their parent galaxy samples in terms of their spectral types, stellar mass, and
star-formation rates/ 24 micron detection.  Cluster galaxies are more likely to be quiescent in the $z-F125W$ v. $F125W-3.6$ $\mu$m plot than
the field sample. More than two-thirds of the cluster mergers and 40\% of the field mergers meet the quiescent galaxy criteria.  
Galaxies classified as `star-forming' are as likely to be detected as 24 micron sources in the cluster than in the field, 
and three of the cluster `quiescent' galaxies have 24 micron emission.     We conclude that at least half, and possibly 2/3 of the cluster mergers
are dissipationless as indicated by their colors and low 24 micron fluxes. 

\begin{figure*}
\plotone{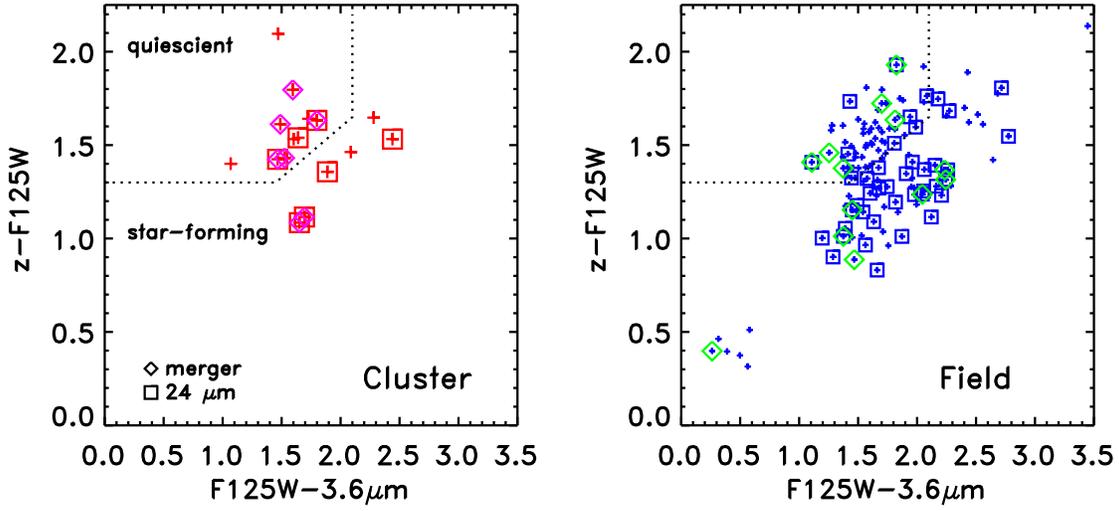}
\caption{ Observed HST F125W - IRAC 3.6$\mu$m  v. z - F125W colors for massive cluster members (left) and control field samples (right) 
with M$_{star} \ge 3 \times 10^{10}$ M$_{\odot}$.  Merger candidates are marked with diamonds. MIPS 24$\mu$m detections are marked with squares.  
Two-thirds of the cluster mergers are in the quiescent region of the plot, while the
remaining third have colors in the star-forming region. }
\end{figure*}

\begin{figure*}
\plotone{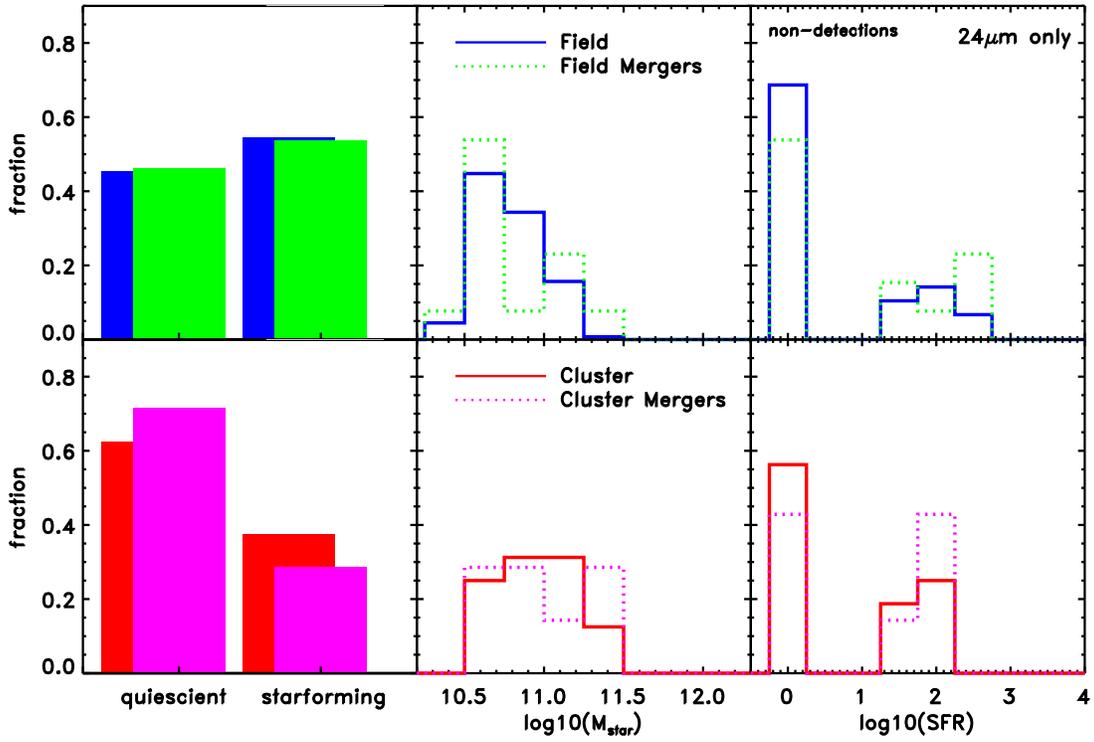}
\caption{The distribution of quiescent/star-forming spectral types (left); stellar mass (center), and 24$\mu$m derived star-formation rates
for field (top) and cluster (bottom) massive galaxy samples. Mergers drawn from the field samples have spectral types, stellar masses, and
SFRs consistent with the overall field population.  Cluster merger properties are consistent with the cluster population, of which $\sim$ 2/3
are quiescent.  } 
\end{figure*}

\section{AGN in Cluster Mergers}
Two of the spectroscopically-confirmed cluster member are point sources in recent Chandra imaging of the proto-cluster (Pierre et al. 2011). 
For these two massive cluster galaxies,  the growth of their central super-massive black holes (SMBH) via accretion and black-hole mergers is coincident
with the assembly of their stellar components.  
CANDELS UDS object 18511 has a [0.5-2 keV] X-ray flux $= 1.60^{+0.36}_{-0.34} \times 10^{-15}$ erg cm$^{-2}$ s$^{-1}$ (Pierre et al. 2011), 
which implies an X-ray luminosity at $z=1.623$  of $L_x$ [0.5-2 keV] $= 1.6 \times 10^{43}$ erg s$^{-1}$, consistent with AGN activity. 
This object has a bright central nucleus and a close secondary nucleus less than 5 kpc away.   CANDELS UDS object 
16582 also has an X-ray AGN with [0.5-2 keV]  X-ray flux $= 2.69^{+0.44}_{-0.42} \times 10^{-15}$ erg cm$^{-2}$ s$^{-1}$, 
corresponding to $L_x$ [0.5-2 keV] $= 2.7 \times 10^{43}$ erg s$^{-1}$ .    
This object has a double nucleus with no clear dominant central component.   Assuming
the AGN activity is associated with one or both stellar nuclei,  then this object is a strong candidate for an offset or dual AGN.
Confirmed dual AGN separated by a few kpc have been observed only in a handful systems, largely at $z \sim 0$ 
( e.g. Comerford et al. 2011; Komossa et al. 2003),  notably in the center of the nearby rich cluster Abell 400 (Hudson et al. 2006). 
An offset AGN is also an intriguing possibility,  as either a signature of a SMBH-SMBH merger in which only one SMBH is visible or
a possible gravitational wave recoiling black hole (e.g. Civano et al. 2010).  A third X-ray AGN is detected in the proto-cluster (Pierre 
et al. 2010),  but this object is not within the CANDELS WFC3 data.  

\section{Implications for massive galaxy assembly}
The implied merger rate for the massive proto-cluster galaxies is extremely high.   For galaxies with double
nuclei and/or projected separations less than $\sim$ 7 kpc physical ($\sim$ 20 kpc co-moving at $z=1.62$) 
and stellar mass ratios $\sim$ 1:1 - 1:10,  the merger detection timescale is $\sim$ 0.20 Gyr (Lotz
et al 2010, 2011).  Adopting a cluster pair fraction between 40-80\% and assuming all of the observed pairs will merge,    
 this gives a merger rate of $\sim 2-4$ mergers per Gyr per galaxy 
for the cluster galaxies, as compared to  $\sim$ 0.2 mergers per Gyr per galaxy in the field.  

Such a high merger rate points to accelerated assembly of the cluster galaxies relative to the field.   Papovich et al. (2011)
finds that the sizes of quiescent cluster galaxies are significantly larger than the $z \sim 1.6$ field population at a fixed stellar
mass,  implying that the cluster early-types have undergone more rapid size growth their recent past. 
The presence of  X-ray AGN in two double-nucleated systems suggests that the SMBH of some $z = 1.62$ cluster galaxies
are growing via accretion and black-hole mergers at the same time as the assembly of their stars.   Yet,  while significant
star-formation and IR activity continues in the cluster (Tran et al. 2010),  the massive cluster galaxies are more likely to have
low star-formation rates and dissipationless mergers than the field.   The recent star-formation histories of the cluster galaxies are 
strongly correlated with their morphologies (Papovich et al. 2011).  Evidence for accelerated evolution in over-dense environments as early as $z \sim 2$ 
has been found by several other recent studies (Natasi et al. 2011; Zirm et al. 2011; 
Cooper et al. 2011).   

We find that XMM-LSS J02182-05102 proto-cluster results are largely consistent with the predictions of de Lucia \& Blaziot (2007)
 for the progenitors of massive clusters galaxies at this epoch. 
The de Lucia \& Blaziot model predicts that the eventual brightest cluster galaxies have only assembled 20\% of
their stars by $z \sim 1.5$.  The most massive cluster galaxies in our sample are $1-2 \times 10^{11}$ $M_{\odot}$, 
consistent with this 20\% value.    However, it is not entirely clear if  massive proto-cluster galaxies can grow quickly enough
to be consistent with being largely assembled by $z \sim 0.8$.  In order for the most massive proto-cluster galaxy at z=1.62  to grow into a 
$\sim 8 \times 10^{11} M_{\odot}$ galaxy by $z \sim 0.8$,  it would need to quadruple its mass. 
This requires at least 1 major merger and $\sim 8-10$ minor mergers in a 2.7 Gyr time period (assuming
$>$ 6:1 mass ratio and $M_{sat} \sim 3 \times 10^{10}$ M$_{\odot}$).
This is on the high end of our observed merger rate of 2-4 mergers per Gyr, and assumes that the merging timescales 
do not evolve significantly either with the growth of the BCG or with the virialization of the cluster. 

The same models also predict that $>$ 90\% of the BGC stars are formed by $z \sim 2$,  well before their assembly into a
single massive galaxy.   
The majority of massive cluster mergers are `quiescent' relative to the full galaxy population at $z \sim 1.6$,  
suggesting that they have already begun to quench their star-formation. Their SEDs and SFRs are reflect the cluster population
as a whole,  which are more likely to be quiescent than field galaxies selected at the same stellar mass ($> 3 \times 10^{10} M_{\sun}$). 
However,  while the majority of cluster members may be fading, they are not completely devoid of star-formation. 
Three of the cluster mergers are detected in 24 micron with implied star-formation rates $> 50$ $M_{\odot}$ yr$^{-1}$ (in the range of 
luminous infrared galaxies).   Therefore, at $z=1.62$ least some proto-cluster mergers are not so-called `dry' mergers and  have 
significant star-formation.   It is unclear if the currently star-forming cluster galaxies will consume their gas reservoirs by $z \sim 1$, or if 
subsequent mergers with gas-rich galaxies can continue to fuel star-formation. On the other hand, the two most massive cluster 
mergers (19085, 18511) show no evidence for 24 micron emission or blue companions, and so may be truly `dry' mergers. 

 It seems unlikely that the high merger rate observed in the center of XMM-LSS  J02182-05102 could continue indefinitely.
We suggest that the virialization of the proto-cluster may act as a mechanism to halt the assembly of the massive cluster galaxies. 
The bulk of the assembly and star-formation of massive cluster galaxies may occur before the virialization of host cluster has completed.  
The massive galaxies will acquire enough mass such that subsequent mergers are likely to have higher mass ratios and therefore
longer dynamical decay timescales.  Galaxies in the local environment will become depleted unless they are replenished by
the accretion of more groups onto the proto-cluster.  The mass growth of the proto-cluster and its eventual virialization will in turn prevent the
efficient accretion of satellites onto the more massive cluster galaxies.   The study of massive galaxies in  a large sample of 
$z > 1.5$ overdensities with a range of cluster masses and virialization states is needed to determine if accelerated galaxy assembly is a generic feature of proto-cluster systems. 

We wish to acknowledge the members of the CANDELS, SEDS, and UKIDSS teams, and M. Cooper for their contributions to the data presented here.   
We thank E. Bell, D. Koo, B. Mobasher,  J.  Newman, L.  Pentericci, A.  Pope,  B. Weiner,  and S. Wuyts for helpful discussions and comments. 
This work is based on observations taken by the CANDELS Multi-Cycle Treasury Program with the NASA/ESA HST, which is operated by 
the Association of Universities for Research in Astronomy, Inc., under NASA contract NAS5-26555.  
This work is supported by HST program number GO-12060. Support for Program number GO-12060 was provided by NASA through a grant from the Space Telescope Science Institute, which is operated by the Association of Universities for Research in Astronomy, Incorporated, under NASA contract NAS5-26555. This work is based on observations made with the Spitzer Space Telescope, which is operated by the Jet Propulsion Laboratory, California Institute of Technology. This work is based in part on data obtained as part of the UKIRT Infrared Deep Sky Survey.

\end{document}